%% file: recsys19-sigconf.tex
\documentclass[sigconf]{acmart}

\usepackage{booktabs} 
\usepackage{multirow}
\usepackage{times}
\usepackage{epsfig}
\usepackage{graphicx}
\usepackage{amsmath}
\usepackage{amssymb}
\usepackage{caption}
\usepackage{subfig}
\usepackage{tabularx}
\usepackage{algorithm}
\usepackage[noend]{algpseudocode}
\usepackage{bchart}
\usepackage{pgfplots}
\usepackage[font=small]{caption}
\usepackage[T1]{fontenc}

\hypersetup{
  colorlinks = true,
  citecolor = green
}

\setcopyright{acmcopyright}
\copyrightyear{2019} 
\acmYear{2019} 
\acmConference[RecSys '19]{Thirteenth ACM Conference on Recommender Systems}{September 16--20, 2019}{Copenhagen, Denmark}
\acmPrice{15.00}
\acmDOI{10.1145/3298689.3347007}
\acmISBN{978-1-4503-6243-6/19/09}

\makeatletter
\renewcommand\@makefnmark{\hbox{\@textsuperscript{\normalfont\color{red}\@thefnmark}}}
\makeatother

\begin{document}
\title{Style Conditioned Recommendations}
\renewcommand{\shorttitle}{Style Conditioned Recommendations}


\author{Murium Iqbal}
\affiliation{%
  \institution{Overstock}
  \city{Midvale}
  \state{Utah}
}
\email{miqbal@overstock.com}

\author{Kamelia Aryafar}
\affiliation{%
  \institution{Overstock}
  \city{Midvale}
  \state{Utah}
}
\email{karyafar@overstock.com}

\author{Timothy Anderton}
\affiliation{%
  \institution{Overstock}
  \city{Midvale}
  \state{Utah}
}
\email{tanderton@overstock.com}

\begin{abstract}
We propose Style Conditioned Recommendations (SCR) and introduce style injection as a method to diversify recommendations. We use Conditional Variational Autoencoder (CVAE) architecture, where both the encoder and decoder are conditioned on a user profile learned from item content data. This allows us to apply style transfer methodologies to the task of recommendations, which we refer to as injection. To enable style injection, user profiles are learned to be interpretable such that they express users' propensities for specific predefined styles. These are learned via label-propagation from a dataset of item content, with limited labeled points. To perform injection, the condition on the encoder is learned while the condition on the decoder is selected per explicit feedback. Explicit feedback can be taken either from a user's response to a style or interest quiz, or from item ratings. In the absence of explicit feedback, the condition at the encoder is applied to the decoder. We show a 12\% improvement on $NDCG@20$ over the traditional VAE based approach and an average 22\% improvement on AUC across all classes for predicting user style profiles against our best performing baseline. After injecting styles we compare the user style profile to the style of the recommendations and show that  injected styles have an average $+133\%$ increase in presence. Our results show that style injection is a powerful method to diversify recommendations while maintaining personal relevance. Our main contribution is an application of a semi-supervised approach that extends item labels to interpretable user profiles.
\end{abstract}

%
%



\maketitle

\input{recsys19-conf}

\bibliographystyle{ACM-Reference-Format}
\bibliography{recsys19-bibliography}

\end{document}

%% file: recsys19-conf.tex
\section{Introduction}

Recommendation systems have a strong popular item bias \cite{abdollahpouri2017controlling}, due to reliance on implicit feedback data. This manifests as the most commonly interacted with items being continually recommended, even when more relevant items are present in the catalog. This can cause users to become stuck in a recommendation filter bubble ~\cite{nguyen2014exploring}, a phenomenon which causes them to be continually exposed to the same types of items, leading to an unpleasant redundant user experience. Item content data can improve recommendations by allowing unpopular items to gain exposure. This data contains faults, though, in the form of false item descriptions or poor quality images. Incorporation of item content data thus risks exposing irrelevant items. Hybrid recommendation systems which make use of item content data and implicit feedback data, are still subject to popular item bias. Using explicit feedback from the user, such as ratings and reviews, can produce results with higher diversity which are more relevant to the user, especially for items which have sparse click data\cite{canamares2018should}. Unfortunately many production recommendation systems cannot leverage this finding as explicit feedback data is limited, requiring manual labeling by users. Our goal is to create a system that can leverage all available data to generate diverse, relevant recommendations which are personalized to a user.

We introduce Style Conditioned Recommendations (SCR) which extends Variational Autoencoder (VAE) recommendations by employing the Conditional VAE (CVAE) architecture. VAE recommendations use implicit feedback data to perform collaborative filtering. We incorporate content data into this approach by introducing a condition on the VAE which is learned from item content data. We refer to this as the user style profile.  In this context style refers to a grouping or genre of interest. In the context of movies, for example, style could be Romance, Comedy, Horror, etc, while for Furniture style could be Modern, Traditional, etc. The model is composed of two parts. The first is an encoder which takes item content data as input, aggregates it into a user content representation and infers user style profiles. We refer to this portion of the network as the text encoder. The second is a CVAE which takes the user item click matrix and the learned user style profiles and generates recommendations. We refer to this as the click VAE.

To leverage explicit feedback data we introduce style injection. This is a novel application of style-transfer techniques, common in computer vision, to the task of recommendations. Styles are injected into recommendations by allowing the user style profile at the encoder to be learned over the data, but selecting the user style profile at the decoder per explicit feedback. This conditions the reconstruction on the new style, and injects it into the recommendations, allowing for incorporation of explicit feedback from a user. Explicit feedback can be incorporated in two ways. The first is by gathering user interests in styles by employing style quizzes, or having users self-identify in interest groups. The second method uses user item ratings. If a user highly rates some items in the catalog, these items can be used to generate a new user profile on which the user recommendations can be conditioned. In the absence of explicit feedback, the user profile can remain unchanged at the decoder, and the system is still able to produce recommendations based only on the implicit feedback and item content data.

To allow for selection of user profiles for injection, we require that user profiles be interpretable. We define interpretability such that each dimension of the user profile represents a pre-defined style. This enables selection as style quizzes can be made to specifically enquire about styles selected for the user profile. The outputs of the quiz can then be directly mapped to a corresponding user profile with which to condition the decoder. We enforce interpretability on the style profiles by employing a label propagation term. User style profiles are learned from item content data, containing limited manually verified style labels. We create a dataset of user content representations with labeled style profiles by sampling the labeled items from our training dataset of user clicks. The result is a dataset which is representative of user click patterns but also bears the style labels.

Throughout this paper we use e-commerce as a case study to frame the problem and examine results. However, it should be noted that the proposed methods are generalizable to all settings that make use of personalized recommendations and are not necessarily limited to e-commerce. Our main contribution is an extension of an unsupervised approach using VAEs to semi-supervised settings. We do so by learning interpretable user style profiles from a limited set of item labels and employing CVAE architecture. This enables our novel style injection approach to recommendations. 

\section{Related Work}
\label{sec:related_work}

Collaborative filtering based recommendations are widely used in industry due to ease of implementation, scalability, high performance and large volume of academic works \cite{linden2003amazon, rendle2009bpr, ning2011slim}. These models work on implicit feedback data to generate recommendations, i.e. they take a user-item click matrix as input. For understanding genres or styles the most popular methods involve topic modeling \cite{hsiao2018creating, hu2014style}. These models often make use of content data to generate recommendations, i.e. they use item text descriptions or item images as input. Hybrid recommendation systems make use of both implicit feedback data and content data \cite{ning2012sparse, al2018adaptive, wang2011collaborative}. These systems provide higher coverage, leveraging the content data for items with low volume of user interactions, but still maintain high accuracy for popular items. We approach hybrid recommendations by learning representations over different modalities of data and concatenating them together as input to a generative process to obtain recommendations \cite{ccano2017hybrid}. This allows us to design the different portions of our network for the specific modality of data they are meant to process. We differ in other works which use this approach in that we learn both representations at the user level to enable style injection. This adds the constraint that representations learned over the content data must be interpretable.

Some personalized recommendation systems aim to learn user profile's in an unsupervised manner from content data \cite{lops2011content}, but these profiles are not directly interpretable. Others build user profiles by incorporating non-platform specific data, such as user locations or information about the user's social network \cite{yin2015modeling, qian2014personalized}. This requires harvesting data from outside the platform to build the user profile, which is not always readily available. Recent work on learning style aware recommendations makes use of visual data to learn style, unsupervised, at the item level \cite{mcauley2015inferring, kang2017visually}. These representations are not directly interpretable either, with even topic modeling based approaches requiring manual interpretation and labeling of the learned topics. Item level style can also be learned to be interpretable with supervised learning \cite{bossard2012apparel, di2013style}, but requires a large corpus of labeled image data. To the best of our knowledge a semi-supervised approach has not yet been applied to learning interpretable style profiles at the user level.

VAEs \cite{kingma2013auto} have recently been applied to the task of generating online recommendations from implicit feedback data (click data) \cite{liang2018variational}. Variational Autoencoders (VAEs) are a natural choice for recommendations as they can rapidly perform collaborative filtering and impute missing values in user item interaction data, one of the most popular recommendation methodologies \cite{portugal2018use}. Further approaches look at incorporating content information, to create hybrid recommendations by using separate VAEs for the content data \cite{li2017collaborative}, but learn representations primarily at the item level, combining item level representations to avoid the cold start problem. Further works look at Conditional VAE (CVAE) architecture for recommendations as well as joint VAE (JVAE) \cite{lee2017augmented}, but does not look at extensions for interpretable encodings. Furthermore, none of these systems make use of style transfer via VAE's for the purpose of recommendations. Style transfer is a technique commonly used in Computer Vision to impose learned styles of one image onto another. This is done in VAEs by conditioning the encoder and decoder on an external label allowing reconstructions to be formed under different selected conditions\cite{makhzani2015adversarial}. As far as we are aware, this methodology has not previously been applied to recommendations. 

Recent work on generating new items from user's preferences has also been studied \cite{kang2017visually, zhu2017your}. In these methods new items or images have been generated that do not necessarily exist on the platform. The methods can therefore be useful for inspiring designers and sellers, but are not immediately impactful for many recommendation systems platforms as the generated images are not of existing items. For example, generating new cover images for movies will not enable further personalization on media streaming platforms, as the generated images won't correspond to an existing item. Our system instead creates new recommendations of existing items based on a user's history and new preferences, allowing the user to explore the catalog of items under different assumed user style profiles.

\section{Method}

In this section we introduce the framework and architecture of SCR as well as the methods which enable style injection. We use the following notation: Bold faced lettering $\mathbf{X}$ represents a matrix, and bold faced lettering with an accompanying arrow, $\Vec{\mathbf{x}}$, represents a vector. Subscript $_T$ indicates content data and related components and encodings, while subscript $_C$ indicates click data and related components and encodings. The letter $z$ represents latent representation and the letters $x$ and $v$ represents raw input. We use a hat to indicate a reconstruction such as $\hat{x}$.

\subsection{Click VAE}
The input to the VAE is a user click matrix $\mathbf{X}_C$ with dimensionality $U \times I$. $U$ is the number of users represented in the dataset and $I$ is the number of items. The model learns compressed latent representations, $\mathbf{Z}_C$, of the input and uses these to reconstruct the input matrix and impute missing values. Recommendations are then obtained by taking all values within the reconstruction which exceed a threshold as recommendations, or by taking the top-N items by value per row in the reconstruction as recommendations for the respective users. 

The distribution over the latent representations, $\mathbf{Z}_C$ must be selected such that the probability of our observations, $p(\mathbf{X}_C)$ also known as the evidence, is maximized. This requires us to compute the posterior, $p(\mathbf{Z}_C|\mathbf{X}_C)$, so that the distribution of the latent variable is conditioned on the observations, allowing us to select the most relevant configuration for $p(\mathbf{Z}_C)$. This in turn requires calculation of $p(\mathbf{X}_C)$ itself, which is intractable as marginalizing over all possible configurations of the latent variable $\mathbf{Z}_C$ is prohibitively expensive. As direct calculation of the true posterior, $p(\mathbf{Z}_C|\mathbf{X}_C)$ distribution, is intractable, an approximation to the true posterior is used $q(\mathbf{Z}_C|\mathbf{X}_C)$. To ensure that this approximation follows our assumed distribution over the posterior $p(\mathbf{Z}_C|\mathbf{X}_C)$, minimizing the Kullback-Liebler divergence, $KL$, between the two distributions is desired. 

\begin{equation}
\begin{aligned}
            & KL(q(\mathbf{Z}_C|\mathbf{X}_C) || p(\mathbf{Z}_C|\mathbf{X}_C)) = {} \\
            & E[\mathbf{log}q(\mathbf{Z}_C|\mathbf{X}_C)] - E[\mathbf{log}p(\mathbf{X}_C, \mathbf{Z}_C)] + \mathbf{log}p(\mathbf{X}_C) \\
\end{aligned}
\end{equation}

Rearranging these terms can lead us to an objective function, known as the Evidence Lower BOund (ELBO), which is necessarily less than the value of $p(\mathbf{X}_C)$, as the KL term is always non-negative. Maximizing the ELBO will allow us to maximize $p(\mathbf{X}_C)$ without directly calculating it. We define the ELBO as:

\begin{equation}
\begin{aligned}
\mathbf{ELBO} = {} & E[\mathbf{log}p(\mathbf{X}_C, \mathbf{Z}_C)] - E[\mathbf{log}q(\mathbf{Z}_C|\mathbf{X}_C)] \\
            & = E[\mathbf{log}p(\mathbf{X}_C, \mathbf{Z}_C)] - KL(q(\mathbf{Z}_C|\mathbf{X}_C) || p(\mathbf{Z}_C)) \\ 
\end{aligned}
\end{equation}

The two terms in the ELBO can be seen as a reconstruction loss, and a regularization, enforcing some prior, $p(\mathbf{Z}_C)$ on the latent representations. For the purposes of recommendations, a multinomial is assumed at the output of the generator \cite{liang2018variational}. This enables a list-wise approach for the recommendations, as the items must compete for limited probability mass, preventing the model from giving over-confident results. As such, this first term is taken as a cross-entropy over the softmax, $\sigma$ of the outputs. The second term is taken as the KL divergence between the approximate multivariate Gaussian, and a standard normal Gaussian prior assumed as the true distribution of the latent variables $p(\mathbf{Z}_C)$.

\begin{equation}
\begin{aligned}
\mathcal{L} = {} & \sum ( \mathbf{X}_C \times \mathbf{log} (\hat{\mathbf{X}_C} ))  \\
            & - 0.5 \times (\mathbf{\mu}^2 - \mathbf{I} - \mathbf{log}\mathbf{det}(\mathbf{\Sigma}) + \mathbf{tr}(\mathbf{\Sigma}))
\end{aligned}
\end{equation}

In VAEs, we parameterize the inference distribution and the generative distribution by an encoder and decoder respectively. To allow for sampling of the latent user click representation, while still allowing for back propagation, the reparameterziation trick is applied \cite{kingma2013auto}; as such the encoder provides the parameters which dictate the distributions over each element of $\mathbf{Z}_C$. These are mean, $\mathbf{\mu}$ and variance, $\mathbf{\Sigma}$. Sampling over a normal Gaussian is done outside of the network, and provided as another input, $\mathbf{\epsilon}$. The terms are then combined to obtain the latent representations.

\begin{equation}
\begin{aligned}
\epsilon \sim \textit{N} (0,1) \\
\mathbf{Z}_C = \mathbf{\mu} + \mathbf{\Sigma}^{\frac{1}{2}} \circ \mathbf{\epsilon}
\end{aligned}
\end{equation}

$\circ$ represents the Hadamard product.

\begin{figure}[t!]
\centering
        \includegraphics[scale=0.2,clip]{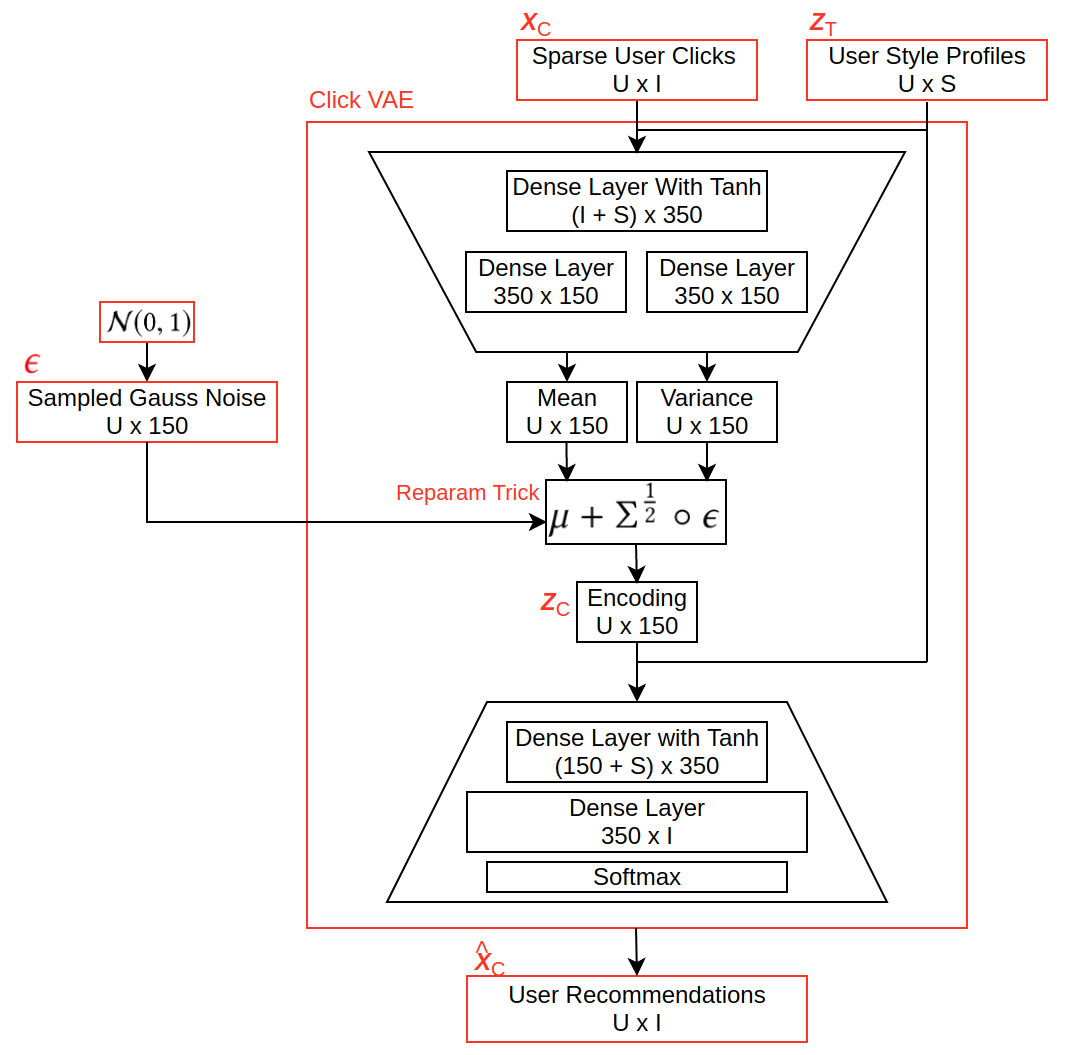}
\vspace{1mm}
\caption{A diagram of the click vae of our model. The input is a sparse matrix of user item clicks and the learned user profile as output by the text encoder. The network uses $tanh$ activations between layers and a $softmax$ activation at the output, which allows for a list-wise approach to recommendations where each item has to compete with others for limited probability mass}
\label{fig:click_vae}
\end{figure}

This latent representation is provided as input to the decoder which produces recommendations. Our system furthers this VAE approach by conditioning both the distribution of clicks, and their latent representations on a user profile, $\mathbf{Z}_T$. We will refer to the VAE portion of our network as the click VAE, a diagram of which is provided in Figure \ref{fig:click_vae}. The updated ELBO reflecting the conditioning is expressed as

\begin{equation}
\label{eq:ELBO}
\begin{aligned}
\mathbf{ELBO} = {} & E[\mathbf{log}p(\mathbf{X}_C, \mathbf{Z}_C, \mathbf{Z}_T)] - KL(q(\mathbf{Z}_C|\mathbf{X}_C, \mathbf{Z}_T) || p(\mathbf{Z}_C)) \\ 
\end{aligned}
\end{equation}

\begin{figure}[t!]
\centering
        \includegraphics[scale=0.2,clip]{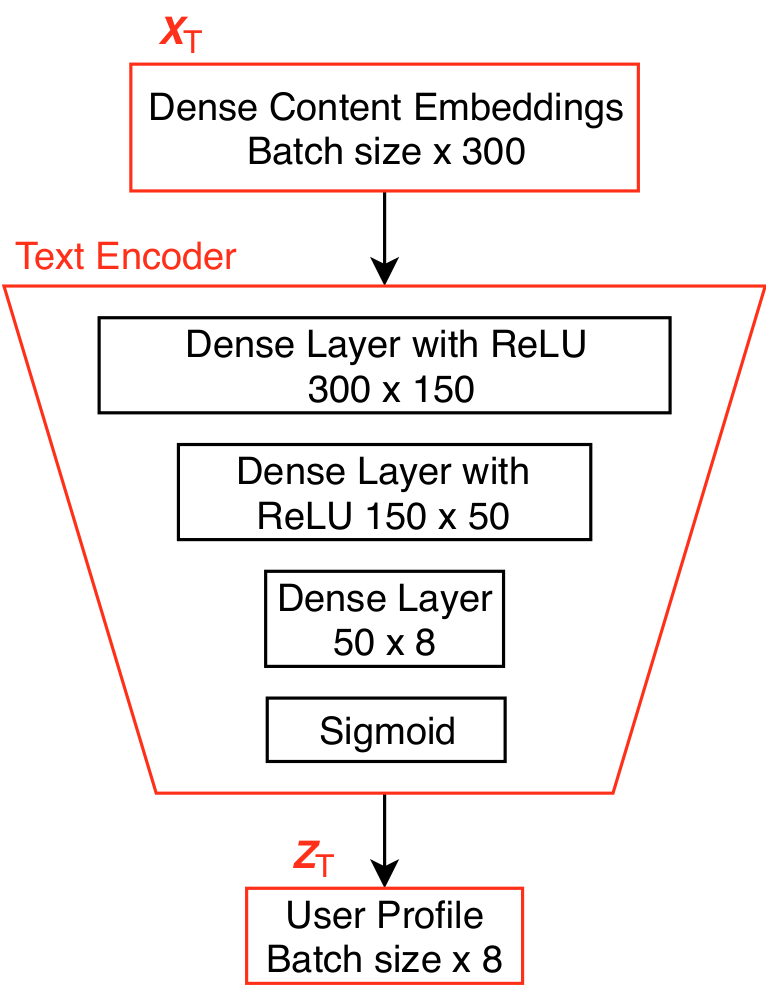}
\vspace{1mm}
\caption{A diagram of the text encoder of our model. The input is a dense matrix of content embeddings. The network uses $ReLU$ activations between layers and a sigmoid activation at the output. The learned user profile represents the probability of the user's interest in each style.}
\label{fig:text_encoder}
\end{figure}

\subsection{Text Encoder}

This is conditioning achieved in the VAE by simply concatenating the user profile $\mathbf{Z}_T$ row-wise to both the input of the encoder $\mathbf{X}_C$ and the input of the decoder $\mathbf{Z}_C$. This user profile is learned from a representation of content data, $\mathbf{X}_T$. The user's content representation is fed into a multi-layer perceptron, similar to the encoder of the click VAE, to obtain the user profiles, $\mathbf{Z_T}$. We will refer to this encoder as the text encoder, a diagram of which is provided in Figure \ref{fig:text_encoder} depicting size and structure of the network.

Users' content representations are obtained by averaging over the content representation of the items with which they have interacted, $\mathbf{M}_T$. Item content data can be provided in any vector form such as embeddings of item text or item images. Our dataset contains item representations built on item text. Item text is taken as item name and item attributes, which are short form tags applied to items describing aspects such as color, material, or size. This text is stripped of stop words, stemmed and tokenized. Each token is then passed through a pre-trained word2vec \cite{mikolov2013distributed} model to obtain a word embedding. The item representation is taken as the average of these embeddings.

\begin{algorithm}[b!]
\caption{User content representation}
\begin{algorithmic}[1]
\Procedure{Get Item Vector}{item Text}
\State bag of words $\gets$ \textbf{tokenize} (item Text)
\State bag of words $\gets$ \textbf{stem} (word) $\forall$ word $\in$ bag of words
\State ${\mathbf{W}} \gets ${} \textbf{word2vec}(word) $\forall$ word $\in$ bag of words
\State $\Vec{\mathbf{m}_T} \gets $ \textbf{mean}($\mathbf{W}$, axis=0)
\State \textbf{return} ${\Vec{\mathbf{m}_T}}$
\EndProcedure
\Procedure{Get User Text Vector}{$\textbf{ User Clicks}$ }
\State items $\gets \binom{n}{k}$ UserClicks
\State $\mathbf{M} \gets ${} Get Item Vector(item) $\forall$ item $\in$ items
\State ${\Vec{\mathbf{x}_T}} \gets \mathbf{mean}$($\mathbf{M}$, axis=0)
\State \textbf{return} ${\Vec{x_T}}$
\EndProcedure
\end{algorithmic}
\label{alg:usercontent}
\end{algorithm}

\begin{algorithm}[b!]
\caption{Create Text Encoder Training Data}
\begin{algorithmic}[1]
\State $\mathbf{V}_T \gets$ []
\State $\mathbf{S} \gets$ []
\For {\texttt{i = 1...10}}
\For {\texttt{u $\in$ Users}}
\State Clicks $\gets$ UserClicks $\cap$ ItemLabels
\State items $\gets \binom{n}{k}$ Clicks
\State ${\textbf{T}} \gets$ \textbf{Get Item Vector}(item) $\forall$ item $\in$ items
\State ${\Vec{\textbf{v}_T}} \gets$ \textbf{mean}($\textbf{T}$, axis=0) 
\State ${\Vec{\textbf{s}}}$ $\gets$ \textbf{mean}(Item Label) $\forall$item $\in$ items 
\State ${\Vec{s}} \gets$ [\textbf{1} (if i > $\theta$) else \textbf{0} $\forall$ i $\in \Vec{\textbf{s}}]$; $\theta = \frac{1}{k}$
\State $\textbf{V}_T \gets [\textbf{V}_T, \Vec{\textbf{x}_T}]$
\State $\textbf{S} \gets [\textbf{S}, \Vec{\textbf{s}}]$
\EndFor
\EndFor
\State Text Encoder $\gets$ \textbf{train Label Prop}($\textbf{V}_T$, S)
\end{algorithmic}
\label{alg:textencodertraining}
\end{algorithm}

\begin{figure}[t]
\centering
 
\subfloat[User click data is used to generate a sparse multi-class vector, $\Vec{\mathbf{x}}_C$. Items are sampled. Their text is used to produce content data representations from a word2vec. These are averaged to obtain the user content vector, $\mathbf{x}_T$.]{
	\includegraphics[width=0.5\textwidth]{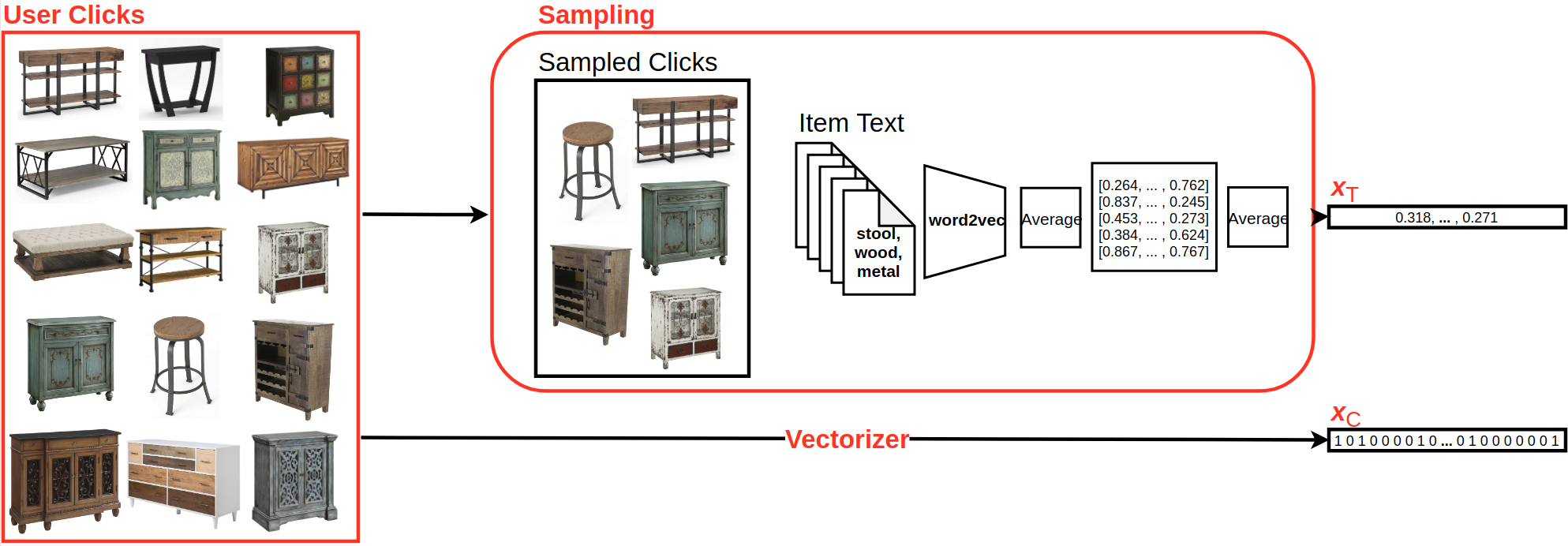} } 
 
\subfloat[The prepped user data is passed through the SCR. The output of the text encoder is used as input at both the click VAE's encoder and decoder. This manifests as a skip connection.]{
	\includegraphics[width=0.5\textwidth]{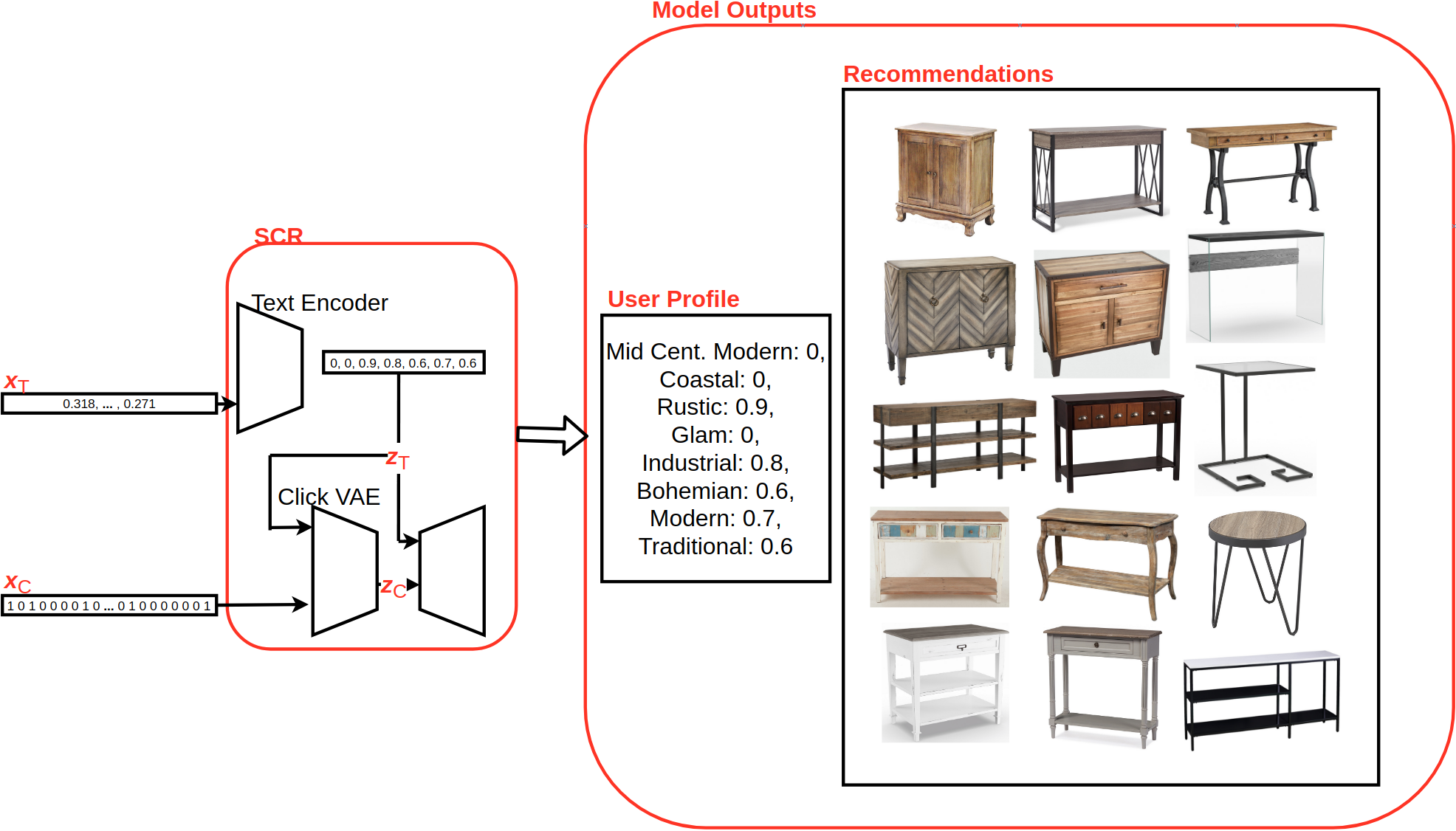} } 
\setlength{\belowcaptionskip}{-10pt}
\vspace{1mm}
\caption{Generation of style conditioned recommendations.}
\label{fig:system_diagram}
 
\end{figure}

Taking an average over all items a user has interacted with yields flatter distributions for users with many interactions than those with fewer interactions. Though the majority of users have few item interactions, the distribution has a long heavy tail, containing users with more than ten times the average number of interactions. This causes distributions dependent on averaging to shift depending on the length of browsing history of a user. To avoid this issue, all users are taken as the average of only a fixed small number $k$ of items which they have interacted with. These items are selected at random each epoch, where $k < N$, where $N$ is the minimum number of items each user has clicked. The process to generate the user content vectors is documented in Algorithm \ref{alg:usercontent}. The result from this process is a $U \times D$ matrix, where $U$ is the number of users (the same as for the click matrix, $\mathbf{X}_C$) and $D$ is the dimensionality of the embedding space. This matrix is then used as input for the text encoder, built of a simple MLP structure with 2 hidden layers. The resulting encoding, $\mathbf{Z}_T$ is the user profile on which the click VAE is conditioned. A diagram of the process to sample and create recommendations with SCR is provided in Figure \ref{fig:system_diagram}.

\subsection{Shaping User Profiles}
User profiles can be created to be interpretable, indicating interest in genres, styles, or color palettes. We define interpretable user profiles such that each dimension in the latent space of the profiles indicates a user's probability of interacting with items of a specific type.  Interpretable profiles enable style transfer techniques, common in computer vision, to be applied to recommendations. Style transfer via a conditional VAE is done by updating the condition at the decoder, allowing the structure captured by the encoding to be transferred to a different condition \cite{makhzani2015adversarial}. In recommendations, this allows user-item preferences, captured in the encoding, to be transferred from one taxonomy of items to another, or from one genre of items to another, from one color palette to another, etc. This would manifest at the decoder as recommendations which are relevant to the customer based on their prior browsing history but which also contain items indicative of the new interest. For this reason we refer to this as "injection". 

Although users are not labeled with style profiles in our dataset, some items have style labels. These labels are limited, with only roughly 2\% of our items having reliable, manually validated labels. To create a dataset which can make use of these labels and enable learning profiles at the user level, we sample the training data, in a similar fashion to that described in Algorithm \ref{alg:usercontent}, but we limit the sampled items to be from those which bear reliable style labels and obtain content representations $\mathbf{V}_T$. We vectorize each item's style label and the corresponding user-level style profile is taken as the average of the labels for the associated items. This average is then thresholded, to obtain a multi-class binary profile, $\mathbf{S}$. The threshold is selected as $\frac{1}{k}$, so that a label of $1$ is only applied to styles which have at least one full items worth of mass in their vector. Not all users within our dataset have interacted with at least $k$ of our labeled items. As such, to create a large diverse corpus of training data, we sample our training set of user to click interactions multiple times to obtain $\mathbf{V}_T$. The label propagation term is then taken as the categorical cross entropy between the true profile $\mathbf{S}$ and the profiles learned from taking $\mathbf{V}_T$ as the input to the text encoder $\mathbf{Z}_V$.

\begin{equation}
\label{eq:label_prop}
\begin{aligned}
\mathbf{Label Prop} = {} & - \sum \mathbf{S} \times \mathbf{log}( \mathbf{Z}_V) \\ 
\end{aligned}
\end{equation}

The process to generate the training data for the label propagation term is documented in Algorithm \ref{alg:textencodertraining}.

\section{Experimentation and Results}
\subsection{Experimental Setup}
Our system is trained on a dataset of user clicks, item content data, and item style labels. We collect a user item click matrix over a span of 3 months. These are filtered to only include users which have interacted with at least 15 items, and items which have had at least 30 users interact with them. This yields a dataset of 177,415 users $\times$ 40,926 items, and a sparsity of 0.1576\%. We have sparse item style labels matrix with dimension 8,136 items by 8 styles. These 8 styles are Mid-century Modern, Coastal, Rustic, Glam, Industrial, Bohemian, Modern and Traditional. Each item can be associated to multiple styles, with 13.7\% of our items being associated with two to three styles, and the rest being associated with just one. Our item content data is taken as the \texttt{word2vec}~\footnote{\url{https://spacy.io/usage/vectors-similarity}} embedding of item title, and item attributes. The $\texttt{word2vec}$ embeddings yield a dense matrix with dimensions 40,926 items $\times$ 300 features.
\begin{table}\centering \scriptsize
\begin{tabular}{@{}|cccccccc|@{}}\toprule
\multicolumn{8}{|c|}{Distribution of Style in Training Data} \\
M-C. & Coast. & Rustic & Glam & Indust. & Boho. & Modr. & Trad.\\ \hline
$0.147$ & $0.048$ & $0.115$ & $0.058$ & $0.043$ & $0.090$ & $0.302$ & $0.170$\\\hline
\end{tabular}
\vspace{1mm}
\caption{Percentage of training samples each style is present in.}
\label{tab:style_training}
\end{table}

To allow for effective training over both loss functions, we freeze part of the network while the other part is actively trained. The datasets for each portion of the network are different, with the text encoder taking in text embeddings, $\mathbf{V}_T$ and style labels, $\mathbf{S}$ for training, and the click VAE taking text embeddings, $\mathbf{X}_T$ and the user click matrix, $\mathbf{X}_C$ as inputs for training. We first train the text encoder to learn the user style profiles by optimizing the label propagation term in Equation \ref{eq:label_prop} while the click VAE portion of the network is frozen. The weights in the text encoder are then frozen, as the click VAE is trained to optimize the ELBO defined in Equation \ref{eq:ELBO}. Adding a scaling factor, $\beta$ to the KL term in the ELBO improves results when $beta < 1$\cite{liang2018variational}. We have empirically chosen to set $\beta = 0.17$.  

\subsection{Datasets}
To validate the performance of SCR on the task of generating recommendations, we take a heldout set of 10,000 users for validation. The users taken for testing and validation have 20\% of their item interactions masked. For validation the unmasked 80\% of clicks are used as input to SCR. The items sampled to create the content representation inputs for SCR are limited to the 80\% of unmasked items as well. The model is scored at how well it can recover the masked 20\% of clicks via normalized discounted cumulative gain (NDCG).

We have a dataset of 8,135 items with style labels. We hold out on-sixth of the items, 1,356 for validation. Sampling to produce the dataset of labeled user style profiles is done only on the training set of user clicks. This allows us to use the validation set of users to examine results of style profiling and style injection. After sampling we have a dataset of 1,573,340 samples for training the user style profiles and samples for validation. The distribution of styles in the training dataset is presented in Table \ref{tab:style_training}.
\begin{table}[H] \small
\begin{tabular}{@{}|c|ccccc|@{}}\toprule
& \multicolumn{2}{c}{NDCG} & \phantom{ab}& \multicolumn{2}{c|}{Recall}\\
& @ 20 & @ 50 && @ 20 & @ 50 \\ \hline
SCR
& $\textbf{0.157}$ & $\textbf{0.192}$ && $\textbf{0.176}$ & $\textbf{0.264}$\\\hline
SCR w/o LP
& $\textbf{0.157}$ & $0.191$ && $\textbf{0.176}$ & $0.263$\\\hline
VAE-CF
& $0.140$ & $0.172$ && $0.155$ & $0.233$\\\hline
cSlim
& $0.095$ & $0.111$ && $0.095$ & $0.137$\\\hline
\end{tabular}
\vspace{1mm}
\caption{Offline evaluation metrics for our proposed model, SCR, against SCR without the label propagation term, VAE-CF and cSLIM. We examine performance of each with NDCG @ 20 and 50. We show that SCR out-performs VAE-CF and cSLIM and prove that addition of the label propagation term, which allows for style profiling and style injection does not detrimentally affect the performance on recommendations.}
\label{tab:ncdg}
\end{table}

\subsection{Performance on Recommendations}
We compare SCR to VAE for collaborative filtering (VAE-CF) \cite{liang2018variational}, and to SLIM \cite{ning2011slim} a form of matrix factorization. We select VAE-CF as our model is an extension upon this recommendation system. We choose SLIM as it has a high performance matrix factorization. As is shown in Table \ref{tab:ncdg}, SCR outperforms both baselines, showing an improvement of $+0.017$ on $NDCG @20$ over the best performing baseline. It should be noted that the addition of the label propagation term does not hurt or improve the results of SCR, with no change on performance at $NDCG @20$. We experimented further with SCR by assuming a Dirichlet prior \cite{srivastava2017autoencoding} over the latent representations $\mathbf{Z}_C$ instead of the standard Gaussian. This did not show an improvement on the task of recommendations. We additionally experiment by training with a Gaussian prior but regularizing the latent space adversarially as in Adversarial Autoencoders (AAE) \cite{makhzani2015adversarial}. This showed an improvement over our results by $+.002$ on $NDCG @20$ but took significantly longer to train, with the Gaussian and Dirichlet priors requiring roughly 60 epochs to converge and the AAE requiring nearly 300. 

\subsection{Performance on User Style Profile Prediction}
\begin{table}[H]\centering \scriptsize
\begin{tabular}{@{}|c|cccccccc|@{}}\toprule
& \multicolumn{8}{c|}{AUC} \\
& M-C. & Coast. & Rustic & Glam & Indust. & Boho. & Modr. & Trad.\\ \hline
SCR
& $0.964$ & $0.946$ & $0.962$ & $0.975$ & $\textbf{0.979}$ & $0.936$ & $0.926$ & $\textbf{0.948}$\\\hline
SCR w/ Gauss
& $\textbf{0.974}$ & $\textbf{0.965}$ & $\textbf{0.965}$ & $0.976$ & $0.975$ & $\textbf{0.950}$ & $0.930$ & $0.946$\\\hline
SCR w/ Dir.
& $0.958$ & $0.953$ & $0.959$ & $\textbf{0.985}$ & $\textbf{0.979}$ & $0.936$ & $\textbf{0.932}$ & $0.940$ \\\hline
LR
& $0.837$ & $0.727$ & $0.812$ & $0.763$ & $0.822$ & $0.729$ & $0.806$ & $0.801$ \\\hline
RF
& $0.671$ & $0.612$ & $0.700$ & $0.684$ & $0.687$ & $0.593$ & $0.667$ & $0.663$\\\hline
\end{tabular}
\vspace{1mm}
\caption{ We show the performance of SCR at predicting user style profiles based on a held-out set of labeled style items. We compare it's performance to that of a mutli-class all-vs-one logistic regression, and multi-class random forest.}
\label{tab:recs_eval}
\end{table}
We compare the performance of SCR to multiclass all-vs-one Logistic Regression. We use per category AUC to evaluate each model's performance on predicting each style. Results for the held-out set are presented in Table \ref{tab:recs_eval}. We select Logistic Regression, as it is similar to the architecture we have chosen for the text encoder, which is a Multi-Layer Perceptron with a sigmoid activation at the output. We chose to compare to a Random Forest classifier as well, as it is a non-linear ensemble model. We allowed for 1000 trees in the Random Forest and no maximum depth defined. Our best performing version of SCR, which assumes a Gaussian prior over the user style profiles improves upon the best performing baseline by $+0.172$ on AUC on average across all styles.

We experiment with the text encoder by also assuming different priors over it. The standard SCR approach assumes no prior. To allow for prior distributions, we perform the same reparameterization trick we use for the click VAE, with the text encoder producing the parameters for the distribution. Without a decoder the loss term only consists of the label propagation term, and a KL term to enforce the assumed prior. We examine the results with a Gaussian prior, and Dirichlet prior. Both showed an improvement to no assumed prior over SCR without any assumed prior over the user style profiles. The Gaussian prior shows an $+0.006$ improvement on AUC on average across all styles over standard SCR. The Dirichlet shows an average $+0.001$ improvement. Results for the two priors are presented in Table \ref{tab:recs_eval}.

\begin{figure}[t!]
\centering
        \includegraphics[scale=0.11,clip]{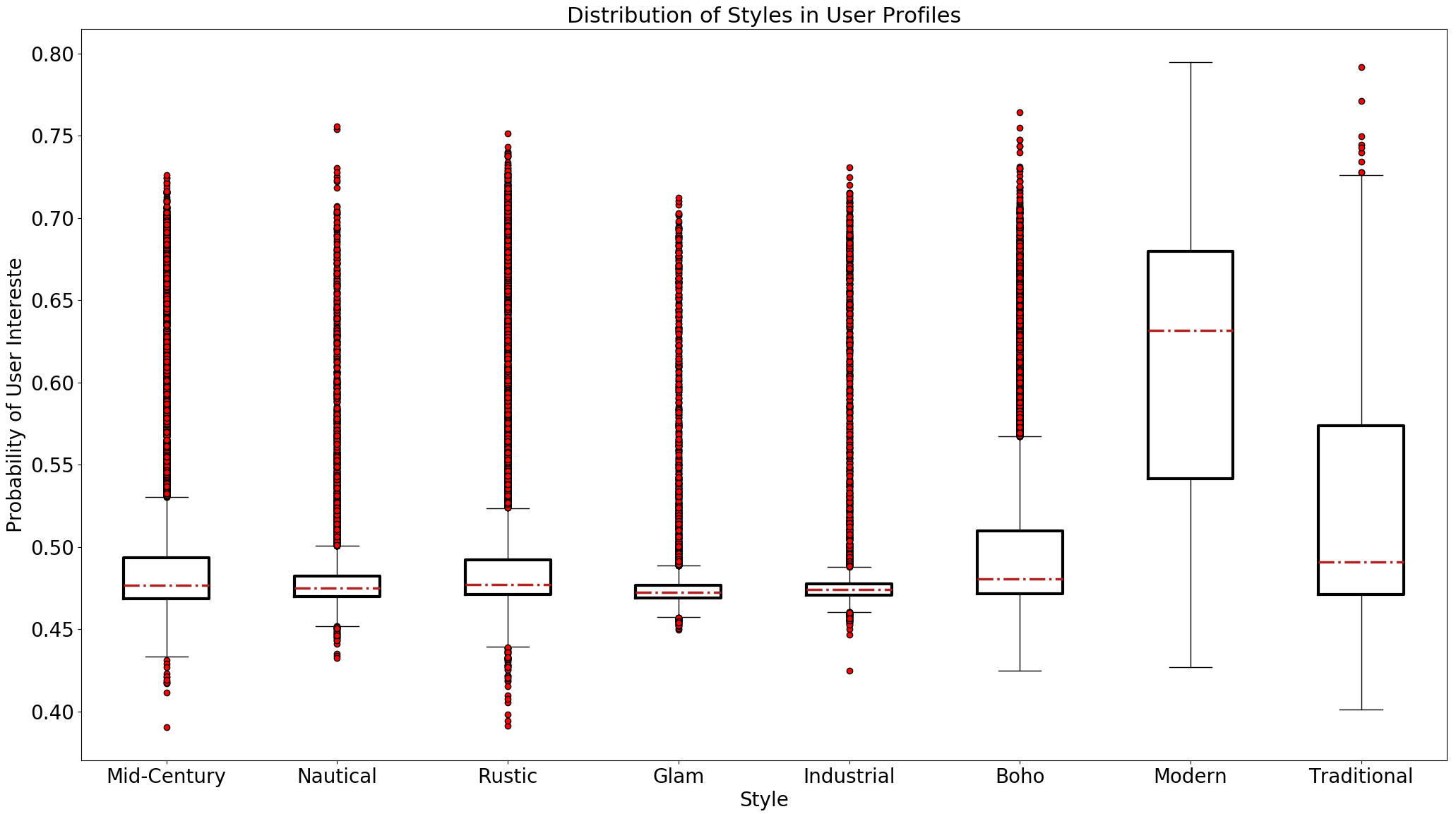}
\vspace{1mm}
\caption{The distribution of styles across user profiles. The Modern style seems to be prevalent across the profiles, while the next most common style is Traditional.}
\label{fig:style_boxplots}
\end{figure}

We examine the distribution of learned style profiles which show a bias towards Modern style and Traditional style as depicted in the distributions of learned user style profiles in Figure \ref{fig:style_boxplots}. This is expected as the highest interacted with items in our dataset are Modern and Traditional items as tabulated in Table \ref{tab:style_training}. We present the correlation between styles in learned user style profiles in Table \ref{tab:style_correlaitons}.

\begin{table}\centering \scriptsize
\begin{tabular}{@{}|c|cccccccc|@{}}\toprule
& \multicolumn{8}{c|}{Pearson Correlation Coeff} \\
& M-C. & Coast. & Rustic & Glam & Indust. & Boho. & Modr. & Trad.\\ \hline
M-C
& $1.0$ & $-0.048$ & $-0.135$ & $-0.053$ & $-0.016$ & $-0.132$ & $-0.142$ & $-0.240$\\\hline
Coast.
& $-0.048$ & $1.0$ & $-0.003$ & $-0.041$ & $-0.043$ & $0.026$ & $-0.243$ & $-0.020$ \\\hline
Rustic
& $-0.135$ & $-0.003$ & $1.0$ & $-0.100$ & $0.132$ & $-0.049$ & $-0.352$ & $-0.032$\\\hline
Glam
& $-0.053$ & $-0.041$ & $-0.100$ & $1.0$ & $-0.022$ & $-0.011$ & $-0.037$ & $-0.071$\\\hline
Indust.
& $-0.016$ & $-0.043$ & $0.132$ & $-0.022$ & $1.0$ & $-0.076$ & $-0.146$ & $-0.100$\\\hline
Boho
& $-0.132$ & $0.026$ & $-0.049$ & $-0.011$ & $-0.076$ & $1.0$ & $-0.298$ & $-0.014$\\\hline
Modr
& $-0.142$ & $-0.243$ & $-0.352$ & $-0.037$ & $-0.146$ & $-0.298$ & $1.0$ & $-0.289$\\\hline
Trad.
& $-0.240$ & $-0.020$ & $-0.032$ & $-0.072$ & $-0.100$ & $-0.014$ & $-0.289$ & $1.0$\\\hline
\bottomrule
\end{tabular}
\vspace{1mm}
\caption{ Displayed are the correlations between different style as measured by the Pearson Correlation Coefficient. All styles are at least weakly negatively correlated with one another, indicating that presence of one style in the profile negatively impacts the likelihood of high interest in another style.}
\label{tab:style_correlaitons}
\end{table}
 
\subsection{Style Injection}
\begin{figure}
\centering
 
\subfloat{
	\includegraphics[width=0.38\textwidth]{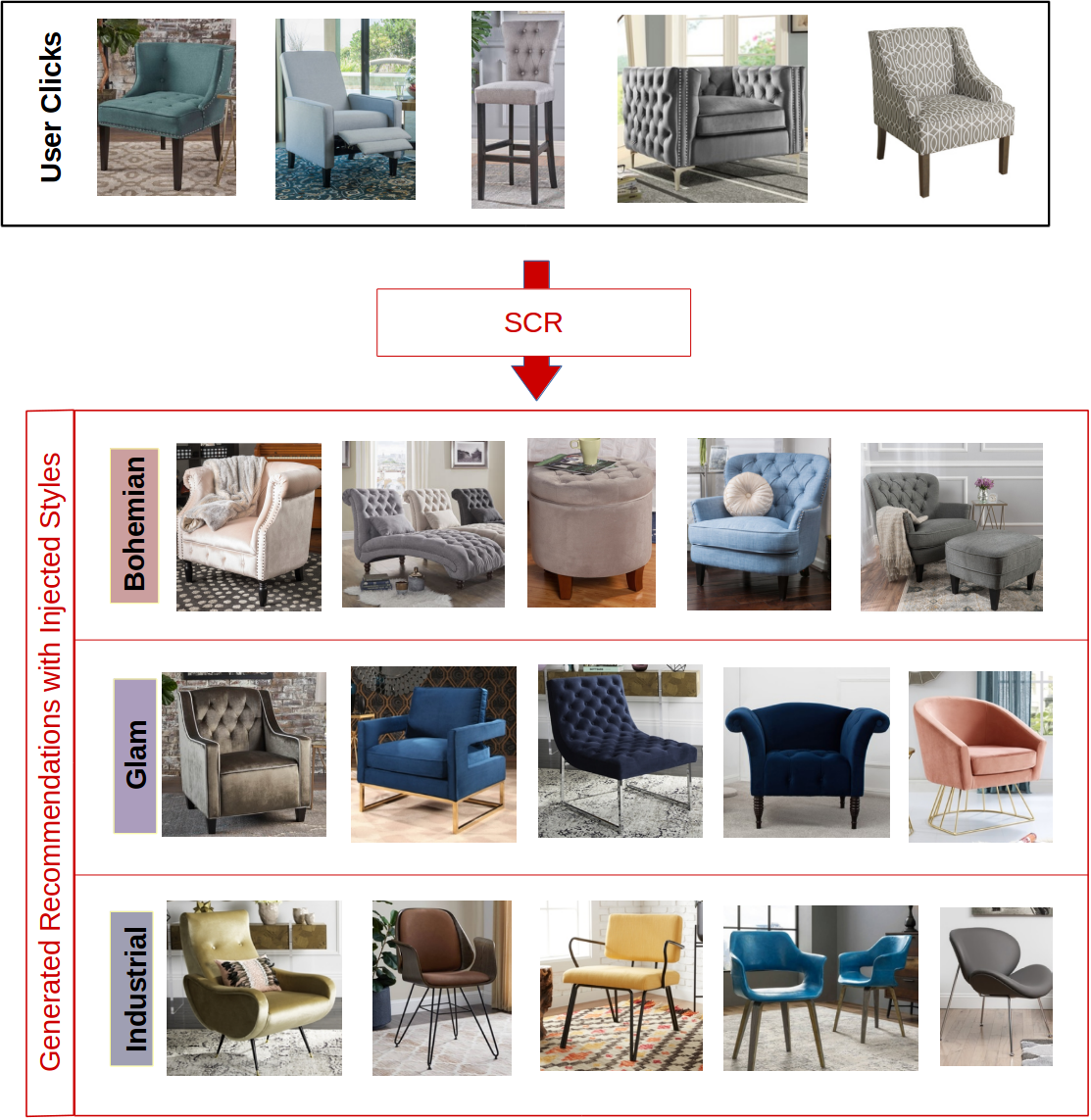} } 
 
\subfloat{
	\includegraphics[width=0.38\textwidth]{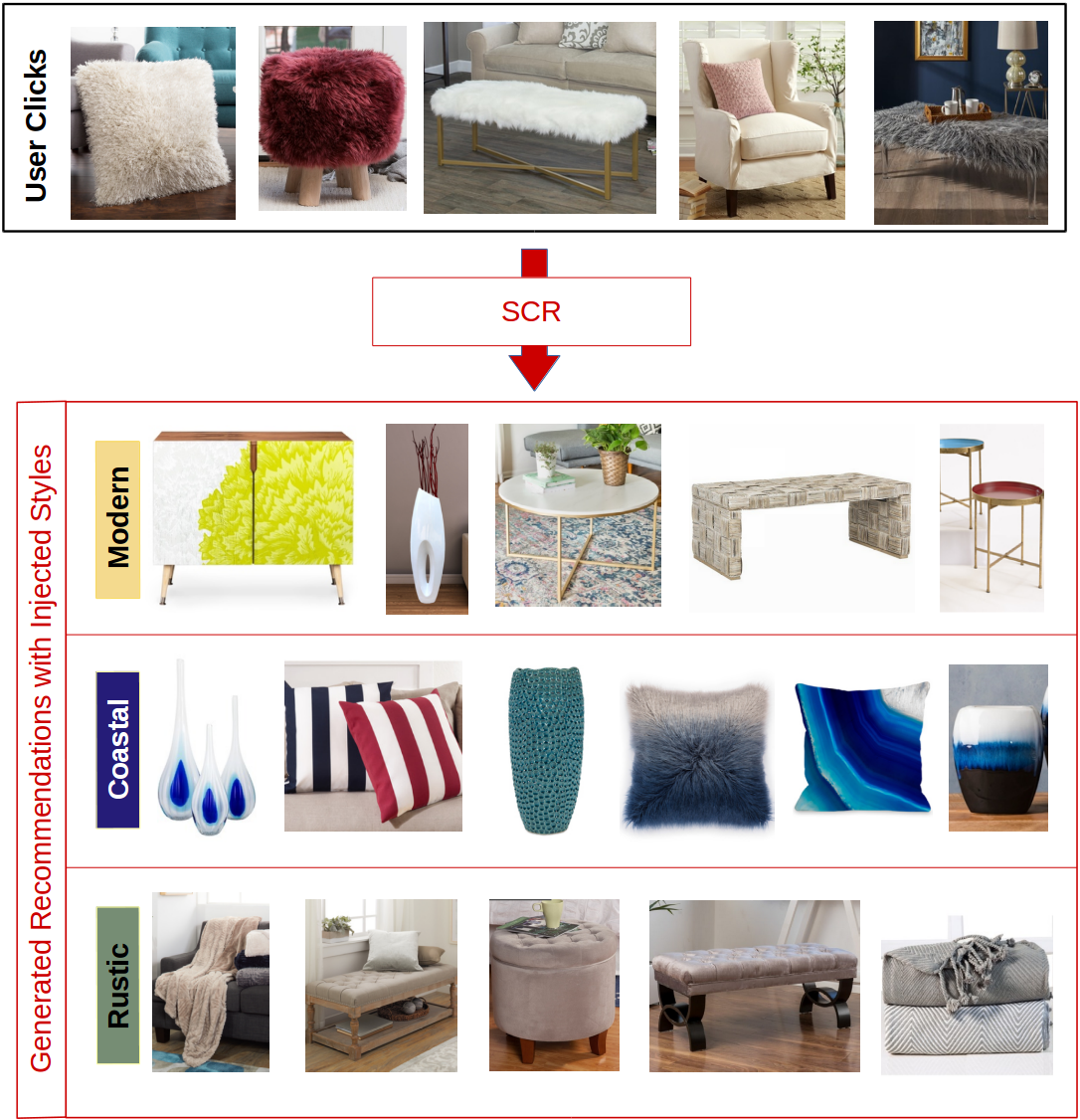} } 
 
\subfloat{
	\includegraphics[width=0.38\textwidth]{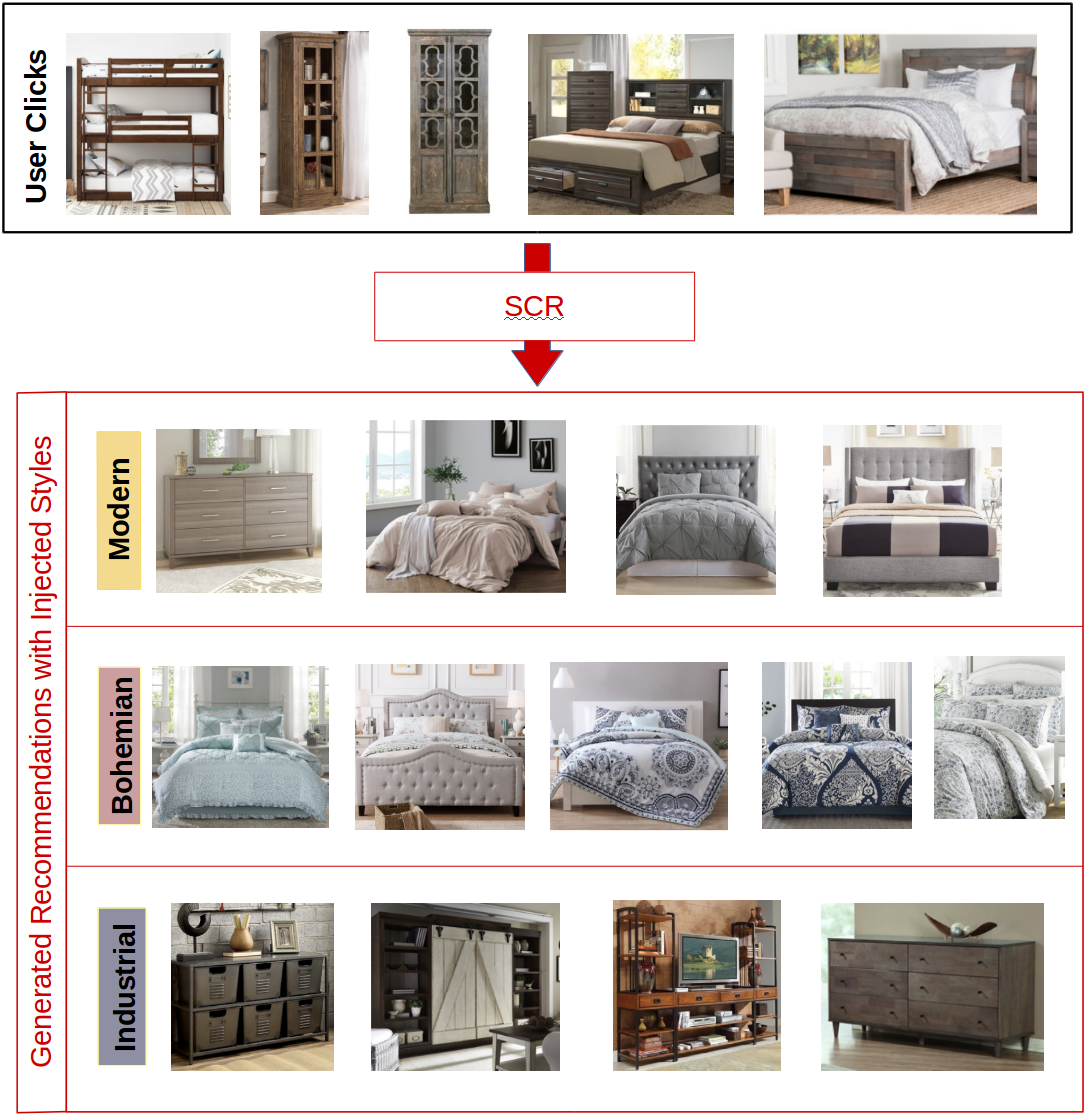}} 
 
\caption{Results of injecting different styles into a user's recommendations.}
\label{fig:style_transfer}
\vspace{1mm}
\end{figure}
\begin{figure}[t!]
\centering
        \includegraphics[scale=0.24,clip]{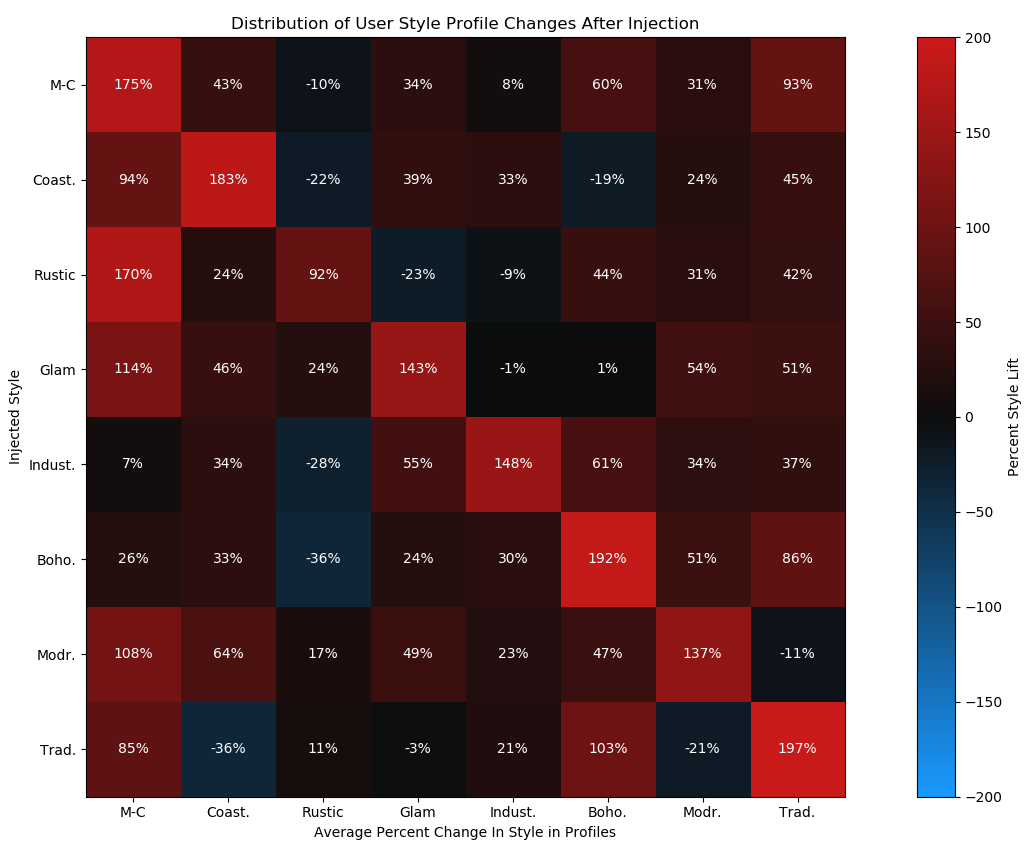}

\caption{The change in style profiles after style injection has been performed. Each row in the figure represents a style being injected, each column shows the average change in that style's value from the user's original style profile.}
\label{fig:style_injection}
\end{figure}

To experiment with style injeciton, users' content vectors $\mathbf{X}_T$ and click vectors $\mathbf{X}_C$ are generated as described previously, and passed through the text encoder and the encoder of the click VAE. At the decoder of the click VAE, the generated $\mathbf{Z}_T$ is replaced with a one-hot encoded vector, where only one of the styles is present within the profile. This allows a user's recommendations to be injected with that style. A visual representation of the style injection in presented in Figure \ref{fig:style_transfer}. To validate whether the style injection is indeed performing as expected, we pass the top 20 recommendations from each injected set into the text encoder, allowing it to sample 5 items at random and produce a new profile. This allows us to infer the style of the recommendations after the injection has been performed. We then subtract each user's original style profile from the style profile learned from the injected recommendations. We display a heat map of the average changes in profiles in Figure \ref{fig:style_injection}. As can be seen, when a style is injected, the dimension corresponding to the injected style sees a large positive shift in the profile.

\section{Discussion}
\label{sec:discussion}
In this section we discuss the results of our experimentation. We examine the affects of architectural decisions on recommendations, the affects of sampling on data distributions and the performance of style injection.

Employing a CVAE architecture for recommendations affords us the flexibility to structure each encoder so that it performs best on the dataset it consumes. The click data, $X_C$ is sparse, while the content data, $X_T$ is dense. We empirically found using $Tanh$ activations for the sparse click data in the click VAE encoder, and ReLU activations for the dense content data provided best results in the text encoder. Having a dropout layer immediately at the input for the text encoder yielded the best results on the style AUC scores, while moving the drop out to the input to the decoder rather than at the input to the entire click VAE yielded the best results for the NDCG validation. Our system is able to use the label propagation term to learn style labels from text data, but any item representation could be substituted. We plan to experiment further with item representations built from images, or both image and text data.

The CVAE architecture and sampling method allows us to incorporate temporal information into the recommendations at inference. The standard VAE approach to recommendations does not incorporate any temporal features; all user clicks are given an equal weight, regardless of how long ago the interaction occured. At training time our system follows this methodology, and ignores any temporal information, sampling uniformly from the items to create the user content representation. At inference time, only the last five items a user has interacted with can be taken to produce the user style profile. This allows recommendations to be conditioned on the user's most recent style profile, making the recommendations highly relevant to the user's recent browsing history while still incorporating their entire history.

\begin{figure}[t!]
\centering
        \includegraphics[scale=0.15,clip]{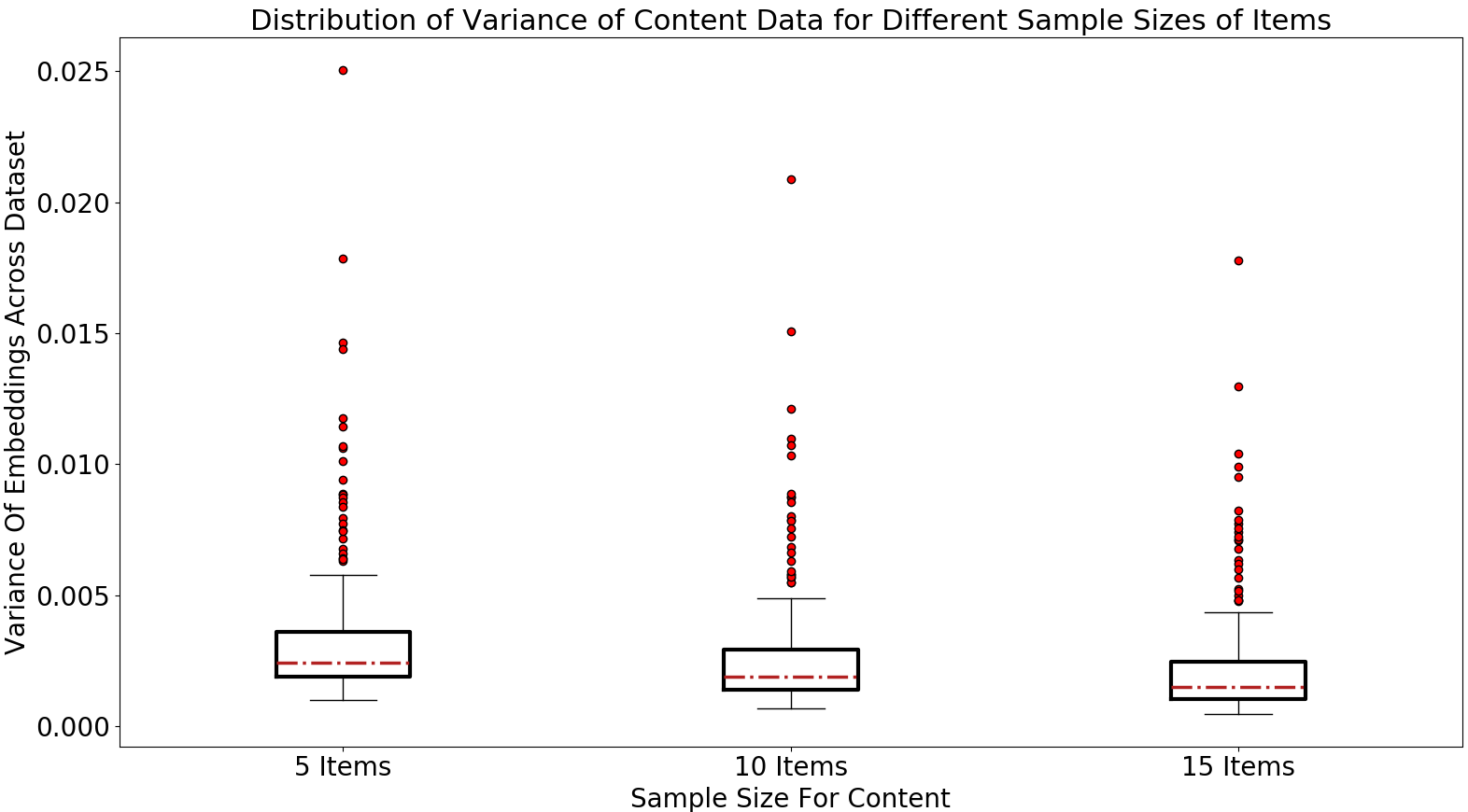}
\vspace{1mm}
\caption{The variance of each feature in the content representation changes with larger sample sizes of items, showing less variance the larger the size. As such we choose a sample size of 5, with enough information to capture a style, but not so large a sample to lose variance in the underlying data once the item embeddings are averaged.}
\label{fig:boxplots}
\end{figure}

Sampling the items used to construct the user content data, instead of using all the items, allows us to ensure the distributions provided for all users are similar. Figure \ref{fig:boxplots} shows that the variance in the feature space of the embeddings diminishes as the number of items used to obtain the average is increased, even though the number of users represented remains the same. Using just $5$ items affords us high variance, while still taking a representative sample of items. This ensures that the distribution user content data is consistent across users. Furthermore, at inference time, accessing the full user history of clicks and aggregating all item content data may be time consuming. Limiting content representations to only depend on $5$ items enables online recommendations by preventing lengthy aggregation computations.

We would like to argue that the effectiveness of the model at predicting style is due to the fact that we are modeling style at the user level rather than the item level. Style is more obvious for groups of items rather than a single time. The same jacket can be worn with business attire to seem formal, but also with casual clothing to be more dressed-down. While the style of the jacket may be ambiguous and subjective, the style of each outfit is distinct. Thus learning style at the user level affords high performance, as shown in our results.

Thresholding the style labels to produce a multi-label one hot output has allowed us to obtain high performance on predicting user style profiles. This approaches assumes a multivariate bernoulli distribution for each labeled profile, and attempts to minimize a categorical cross-entropy between the labels and the learned profiles. This allows multiple styles to be present in the same profile. We experimented with assuming a multinomial distribution over the user style profiles. No thresholding was applied and the model attempted to minimize a mean-squared error (mse) loss term. This yielded poor results, with the model continually converging to a uniform distribution of predicted values across styles for all users. Assuming a multi-variate bernoulli enabled the network to infer distinct profiles per user. The correlations between styles, as shown in Table \ref{tab:style_correlaitons} shows that even though we did not assume a multinomial distribution over the profiles, which would cause each style to compete with others for the limited probability mass, the learned styles still show a negative correlation with one another. This shows that the model has learned that the presence of one style negatively impacts the likelihood of the presence of other styles.

Style profiles learned under the Gaussian prior perform best. This could be due to the reparametrization trick, which introduces a form of noise based regularization. This allows the model to generalize the learned representations better. The Dirichlet prior performs worse than Gaussian, because it assumes a multinomial distribution over the user style profile. This causes the model to perform poorly as the labels assume a multi-variate bernoulli distribution. Pairing the Dirichlet prior with the mse loss term still suffered from a convergence to uniform distributions over the user style profiles.

For styles which are negatively correlated, the model does not succeed at injecting the one into a user's recommendations if they are already aligned with the other. We have noticed in these cases, the system gracefully falls back to the user's original recommendations. We think this is preferred behavior over exposing items which are unrelated to the user but are relevant to the requested style.

\section{Conclusion}
\label{sec:conclusion}
We introduce style injection via Style Conditioned Recommendations (SCR) as a novel way to diversify recommendations while maintaining relevance to users. Explicit feedback, though hard to gather, gives a clear picture of the user's interests and tastes. Implicit feedback allows recommendation systems to leverage patterns of shopping behavior across users to find other relevant items, but creates a heavy popular item bias. Item content data allows all items to be considered, but is subject to faults such as false item descriptions, missing or poor quality item images, poorly named items, and more. SCR is able to leverage all three of these signals to produce relevant personalized recommendations. By employing semi-supervised learning, we leverage a limited number of cleanly labeled items to learn user level style profiles. SCR affords increased diversity as explicit feedback becomes available by using style injection. 

In the absence of explicit feedback for style injection, SCR is still a performant recommendation model. SCR is able to infer style profiles and generate recommendations. It outperforms both VAE-CR method \cite{liang2018variational} and cSLIM \cite{ning2011slim} on the task of recommendations and out performs baselines of multi-class all-vs-one logistic regression and multi-class random forest on predicting user style profiles.